\documentclass[twocolumn,showpacs,prb,amsmath,amssymb]{revtex4}
\usepackage{graphicx}
\usepackage{bm}

\DeclareMathOperator{\Ai}{Ai}
\begin{document}
\title{Gate control of spin dynamics in III--V
semiconductor quantum dots}
\author{Rogerio \surname{de Sousa}}
\author{S. \surname{Das Sarma}}
\affiliation{
Condensed Matter Theory Center, Department of Physics,
University of Maryland, College Park, MD 20742-4111}
\date{\today}
\begin{abstract}
We show that the $g$-factor and the spin-flip time $T_{1}$ of a
heterojunction quantum dot is very sensitive to the band-bending
interface electric field even in the absence of wave function
penetration into the barrier.  When this electric field is of the
order of $10^{5}$~V/cm, $g$ and $T_{1}$ show high sensitivity to
dot radius and magnetic field arising from the interplay between
Rashba and Dresselhaus spin-orbit interactions.  This result opens
new possibilities for the design of a quantum dot spin quantum
computer where $g$-factor and $T_{1}$ can be engineered by
manipulating the spin-orbit coupling through external gates.
\end{abstract}
\maketitle
%

Understanding and controlling the behavior of spins in
semiconductor heterostructures may lead to a whole new class of
devices,\cite{dassarma01,recher00} ranging from spin-polarized p-n
junctions\cite{zutic02} to a quantum dot spin quantum
computer.\cite{loss98} Electrical control over spin-orbit coupling
parameters in a quantum well has long been suggested as an
effective means to manipulate spin,\cite{datta90} and was recently
experimentally demonstrated in InAs
heterostructures.\cite{nitta97} This was possible because the
Rashba\cite{bychkov84,silva97} and the
Dresselhaus\cite{dresselhaus55} spin-orbit interactions are
sensitive to the electric field providing vertical confinement to
a two dimensional electron gas (2DEG), which is approximately
proportional to the 2DEG density and can be controlled through a
gate voltage. When additional gates provide lateral confinement, a
single electron can be trapped in a quantum dot
(QD),\cite{tarucha96} whose orbital states in an external magnetic
field perpendicular to the 2DEG are the well known Fock-Darwin
states.\cite{jacak98} Here we consider the effect of Rashba and
Dresselhaus spin-orbit interactions in the spin-doublet ground
state of a Fock-Darwin QD.  Using an effective mass approximation
and exact diagonalization of a Fock-Darwin subspace we show that
the ground state $g$-factor and spontaneous phonon emission rate
$1/T_{1}$ (due to the spin-orbit admixture
mechanism)\cite{khaetskii01} can be substantially manipulated by
varying the heterojunction electric field in the range of
$10^{5}-10^{6}$~V/cm (corresponding to 2DEG density of the order
of $10^{12} - 10^{13}$~cm$^{-2}$).

Recently, electrical control over GaAs quantum well $g$-factor has
been achieved by forcing the electron wave function to overlap
with the AlGaAs barrier.\cite{salis01,jiang01} Here we
intentionally neglect barrier penetration, to show that overlap
with a different material \emph{is not a necessary condition to
achieve substantial
  electrical control over quantum dot $g$-factor}.  Moreover, by
avoiding barrier penetration one can suppress an additional
spin-lattice relaxation mechanism due to interface
motion.\cite{woods02} These results open new possibilities in the
design of a quantum dot quantum computer: for example, spin qubits can
be brought in and out of resonance to a global spin resonance field by
gate control of their $g$-factor;\cite{vrijen00} a speed up in quantum
computer initialization (setting all spins up for example) can be
achieved by decreasing the spin-flip time $T_{1}$ with a gate voltage.

The Hamiltonian for a single electron bound to an heterojunction
quantum dot can be divided into five parts,
\begin{equation}
{\cal H}= {\cal H}_{\rm{0}}+{\cal H}_{\rm{z}}+{\cal
  H}_{\rm{R}} +{\cal H}_{\rm{D1}}+{\cal H}_{\rm{D2}}.
\label{htotal}
\end{equation}
The first contribution corresponds to a single 2D electron confined in
the $xy$ plane by
a parabolic potential and subject to a magnetic field ${\mathbf B}$,
\begin{equation}
{\cal H}_{\rm{0}}=\frac{{\mathbf P}^{2}}{2m^{*}}
+\frac{1}{2}m^{*}\omega_{0}^{2}r^{2}
+\frac{1}{2}g_{0}\mu_{B}\sigma_{z}B, \label{h0}
\end{equation}
where the kinetic momentum ${\mathbf P}={\mathbf p}+e/c{\mathbf
A}$  is written with the canonical momentum  ${\mathbf
p}=-i\hbar(\partial_{x},\partial_{y}, 0)$ and vector potential
${\mathbf A}=B/2(-y,x,0)$ confined to the 2D plane.  Here $e$ is
the electron charge, $c$ is the velocity of light, $m^{*}$ is the
conduction band edge effective mass, $\omega_{0}$ is the parabolic
confining potential frequency, $g_{0}$ is the Bulk $g$-factor,
$\mu_{B}$ is the Bohr magneton, and $\sigma_{z}$ is the diagonal
Pauli matrix. ${\cal H}_{\rm{0}}$ is diagonal when written as a
function of the Fock-Darwin number operators
$n_{\pm}=a^{\dag}_{\pm}a_{\pm}$,\cite{jacak98}
\begin{equation}
{\cal H}_{\rm{0}}=\hbar \omega_{+}\left(n_{+}+\frac{1}{2}\right)+
\hbar\omega_{-}\left(n_{-}+\frac{1}{2}\right)
+\frac{1}{2}g_{0}\mu_{B}\sigma_{z} B,
\end{equation}
\begin{eqnarray}
a_{\pm}^{\dagger}&=&\frac{1}{2\ell}\left(x\pm i
y\right)-\frac{\ell}{2}\left(\partial_{x}\pm i \partial_{y}\right),
\label{creation}
\\
a_{\pm}&=&\frac{1}{2\ell}\left(x\mp i
y\right)+\frac{\ell}{2}\left(\partial_{x}\mp i \partial_{y}\right).
\label{annihilation}
\end{eqnarray}
Here $\omega_{\pm}=\Omega\pm\omega_{c}/2$, with
$\Omega=\sqrt{\omega_{0}^{2}+\omega_{c}^{2}/4}$ and
$\omega_{c}=eB/m^{*}c$ being the renormalized dot frequency and
cyclotron frequency respectively, with
$\ell=\sqrt{\hbar/m^{*}\Omega}$ being the Fock-Darwin radius which
sets the length scale for the eigenstates
$|n_{+}n_{-}\sigma\rangle$ ($\sigma=\pm 1$ represents the spin
up/down states in the $z$ direction), The second term in the
Hamiltonian (\ref{htotal}) represents the quantum well confinement
in the growth ($z$) direction, ${\cal H}_{\rm{z}} =
p_{z}^{2}/2m^{*}+V(z)$, where $V(z)$ is a triangular well,
$V(z)=eEz$ for $z\geq 0$ and $V(z)=\infty$ for $z<0$.
%
%
\begin{figure}
\includegraphics[width=3in]{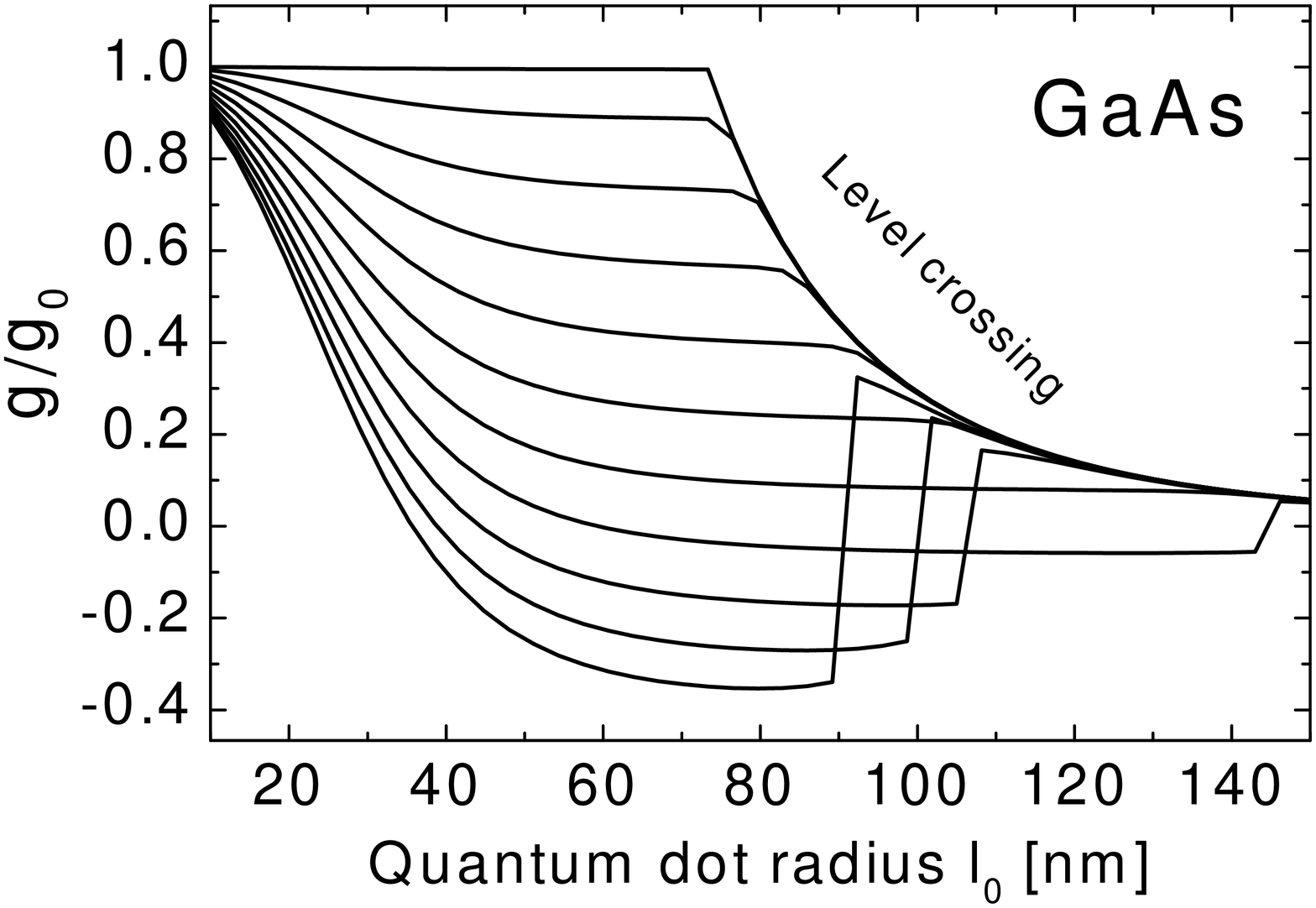}
\caption{Quantum dot $g$-factor divided by the bulk $g$-factor
  as a function of the dot radius $l_{0}=\sqrt{\hbar/m^{*}\omega_{0}}$
  for a GaAs heterojunction. Each curve corresponds to a different
  electric field: From top to bottom, $E=10^{4}$, $1,2,\ldots, 10
  \times 10^{5}$.  The magnetic field applied in the growth [100]
  direction is assumed to be $1$~Tesla. At large $l_{0}$ the two
  lowest energy states have same spin, hence the level crossing. The
  results for GaSb are similar, except that the level crossing occurs
  for smaller $l_{0}$.
\label{figone}}
\end{figure}
A simple numerical calculation leads to the ${\cal H}_{\rm{z}}$
ground state
\begin{equation}
\Psi_{\rm{0z}}(z)= 1.4261 \kappa^{1/2} \Ai{\left( \kappa z +
\zeta_{1}\right)}, \label{groundz}
\end{equation}
where $\zeta_{1}=-2.3381$ is the first zero of the Airy function
$\Ai$, while the inverse length scale $\kappa$ is set by
\begin{equation}
\kappa = ( 2m^{*}eE/\hbar^{2})^{1/3},
\end{equation}
and the ground state energy is $E_{0z}=-\zeta_{1}eE/\kappa$. In
the discussion below we will make use of the average momentum
squared in the state (\ref{groundz}), $\langle p_{z}^{2} \rangle =
0.7794 (\hbar\kappa)^{2}$, and the average position $\langle z
\rangle = 1.5587/\kappa$ (which is the thickness of the 2DEG). We
now turn to the spin-orbit interactions, third to fifth terms in
Eq.~(\ref{htotal}). A  ${\mathbf k}\cdot {\mathbf  p}$ band
structure calculation for zincblende materials\cite{dresselhaus55}
leads to the bulk conduction band spin-orbit interaction
\begin{equation}
{\cal H}_{{\rm Bulk}}=\gamma_{c}/(2\hbar^{3}){\mathbf \sigma}
\cdot\mathbf{\tilde{P}}, \label{bulkso}
\end{equation}
where $\tilde{P}_{x}=P_{x}(P_{y}^{2}-P_{z}^{2})+ \rm{h.c.}$,
$\tilde{P}_{y}$ and $\tilde{P}_{z}$ can be obtained by cyclic
permutations. Notice that Eq.~(\ref{bulkso}) is hermitian and
gauge invariant.  The value of $\gamma_{c}$ is determined by the
band structure parameters of the III--V semiconductors (Table I).
By averaging Eq.~(\ref{bulkso}) over the quantum well ground state
[Eq.~(\ref{groundz})] we get two spin-orbit terms, linear and
cubic in momenta (here the quantum well growth direction is
assumed to be [001]),\cite{dyakonov86}
\begin{eqnarray}
{\cal H}_{\rm{D1}}&=& 0.7794 \gamma_{c} \kappa^{2}/\hbar
\left(-\sigma_{x}P_{x} +\sigma_{y}P_{y} \right), \label{hd1}
\\
{\cal H}_{\rm{D2}}&=& \gamma_{c}/\hbar^{3} \left(
\sigma_{x}P_{x}P_{y}^{2}-\sigma_{y}P_{y}P_{x}^{2}\right)+\rm{h.c.}.
\label{hd2}
\end{eqnarray}
\begin{figure}
\includegraphics[width=3in]{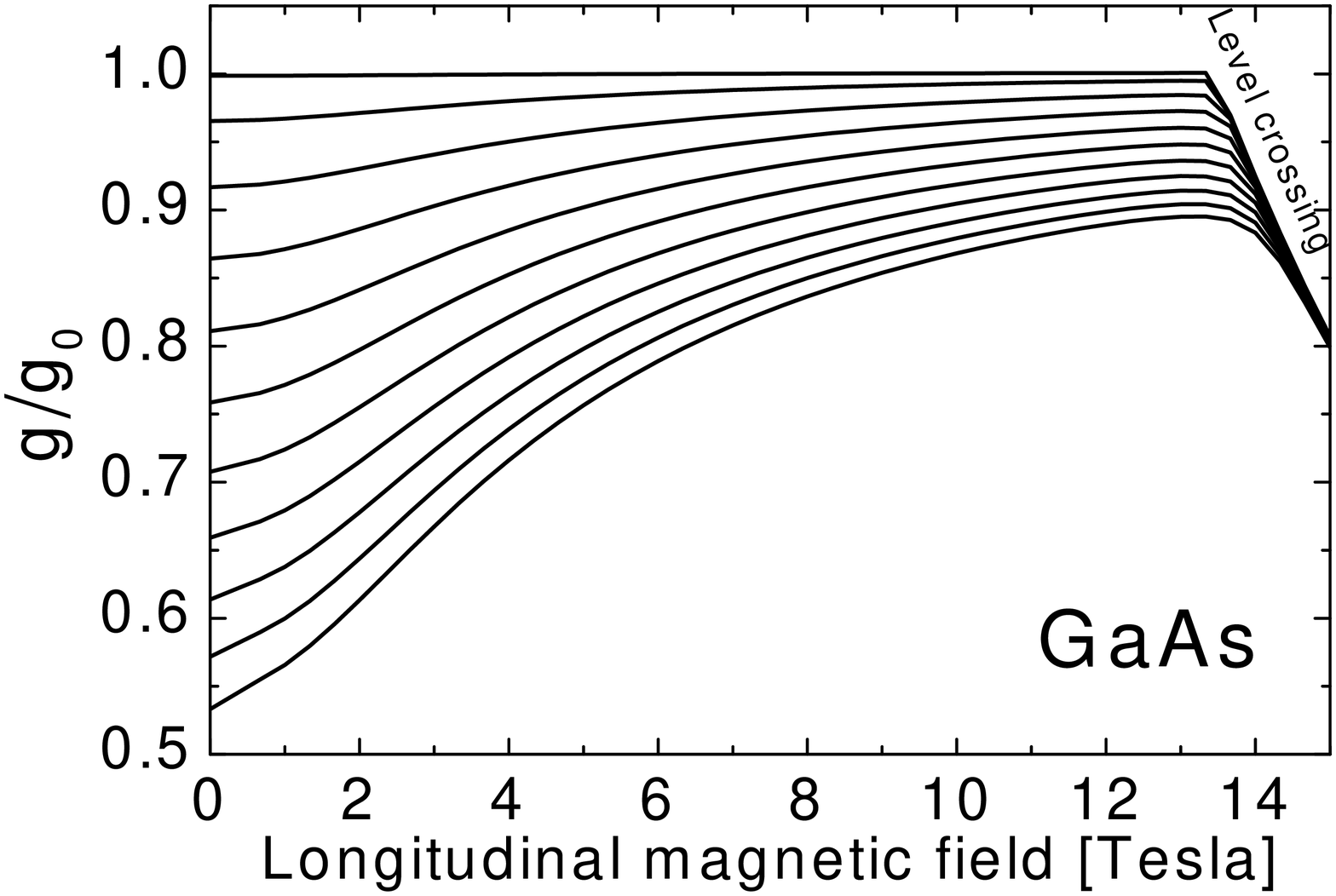}
\caption{Quantum dot $g$-factor as a function of the magnetic
field $B$ for a GaAs heterojunction.  The electric fields are the
same as in Fig.~1.  Dot radius is $l_{0}=20$~nm. At large $B$ a
similar level crossing as in Fig.~1 takes place. \label{figtwo}}
\end{figure}
The structural inversion asymmetry in $V(z)$
leads to the Rashba interaction\cite{silva97}
\begin{equation}
{\cal H}_{\rm{R}}= \alpha_{R}eE/\hbar
\left(\sigma_{x}P_{y}-\sigma_{y}P_{x}\right), \label{hrashba}
\end{equation}
which for the triangular well considered here is directly
proportional to $E$ ($\alpha_{R}$ depends on band structure
parameters).\cite{silva97}  It is useful to condense ${\cal
H}_{\rm{R}}$ and ${\cal H}_{\rm{D1}}$ in a single Hamiltonian
written as a function of the Fock-Darwin operators,
\begin{equation}
{\cal H}_{\rm{D1}}+{\cal H}_{\rm{R}} = \hat{V}
\sigma_{+}+\hat{V}^{\dagger}\sigma_{-}, \label{hd1hr}
\end{equation}
\begin{equation}
\hat{V} = -\alpha_{-}a_{-}^{\dagger}+\alpha_{+}a_{+}
+i \beta_{-}a_{-}-i \beta_{+}a_{+}^{\dagger},
\label{vhat}
\end{equation}
with $\sigma_{\pm}=(\sigma_{x}\pm i \sigma_{y})/2$ and
spin-orbit energy scales defined as
\begin{eqnarray}
\alpha_{\pm} &=& \alpha_{R} e E \xi_{\pm},
\label{alphapm}
\\
\beta_{\pm} &=& 0.78 \gamma_{c}\kappa^{2}\xi_{\pm},
\label{betapm}
\\
\xi_{\pm}&=&\frac{1}{\ell}\pm \frac{eB\ell}{2\hbar c}.
\label{xipm}
\end{eqnarray}
Therefore ${\cal H}_{\rm{R}}$ and ${\cal H}_{\rm{D1}}$ couples
Fock-Darwin levels differing by one quantum number.  The amount of
spin up/down admixture in the ground state is given by the ratios
$\alpha_{\pm}/\hbar\omega_{0}$, $\beta_{\pm}/\hbar\omega_{0}$,
which at $B\approx 0$ are directly proportional to the dot radius
$l_{0}=\sqrt{\hbar/m^{*}\omega_{0}}$. Therefore larger dots will
be more sensitive to spin-orbit coupling, at least within
perturbation theory (see below). Moreover, the question of whether
Rashba or Dresselhaus dominates depends only on material
parameters and electric field, since $\alpha_{\pm}/\beta_{\pm}=
\alpha_{R}/\gamma_{c}(\hbar^{2}/m^{*})^{2/3}(eE)^{1/3}$.  If the
electric field ranges from $10^{4}-10^{6}$~V/cm this ratio equals
$0.1-0.7$ for GaAs, $0.2-1.0$ for GaSb, $1.5-6.8$ for InAs, and
$5.6-26$ for InSb. Therefore for III--V semiconductor quantum dots
it is important to consider the interplay between Rashba and
Dresselhaus spin-orbit interactions. High electric field GaSb
heterojunctions might realize the condition
$\alpha_{\pm}=\beta_{\pm}$, leading to an interesting
simplification of Eq.~(\ref{hd1hr}), which becomes proportional to
$(\sigma_x +\sigma_y)$.\cite{schliemann03} However this symmetry
is broken by Eq.~(\ref{h0}), which contains a magnetic field
pointing in the [001] direction. Hence it will have no
consequences here [the highly symmetric case of
$\bm{B}\parallel$~[110] and $\alpha_{\pm}=\beta_{\pm}$ leads to QD
$g$-factor exactly equal to the bulk value $g_0$, with no
spin-lattice relaxation, as long as the cubic spin-orbit
interaction Eq.~(\ref{hd2}) is neglected].

We now turn to the cubic spin-orbit term [Eq.~(\ref{hd2})],
\begin{eqnarray}
{\cal H}_{\rm{D2}} &=& -i \sigma_{+}\bigglb[
\lambda_{1}a_{-}^{\dag 2}a_{+} + \lambda_{2}a_{+}^{\dag 2}a_{-}
+\lambda_{4}a_{-}n_{-} \nonumber\\&&
-\lambda_{3}a^{\dag}_{+}(n_{+}+1)
+\frac{\lambda_{2}}{3}a_{-}(2n_{+}+1) \nonumber\\&&
-\frac{\lambda_{1}}{3}a^{\dag}_{+}(2n_{-}+1)
+\frac{\lambda_{2}}{3}a^{\dag 2}_{+}a^{\dag}_{-}
-\frac{\lambda_{1}}{3}a_{+}a_{-}^{2} \nonumber\\&&
-\lambda_{4}a^{\dag 3}_{-} +\lambda_{3}a^{3}_{+}\bigglb]+{\rm
h.c.}, \label{hd2fd}
\end{eqnarray}
with $\lambda_{1}=3/4 \gamma_{c} \xi_{+}\xi_{-}^{2}$,
$\lambda_{2}=3/4 \gamma_{c} \xi_{-}\xi_{+}^{2}$, $\lambda_{3}=1/4
\gamma_{c} \xi_{+}^{3}$, $\lambda_{4}=1/4 \gamma_{c} \xi_{-}^{3}$.
${\cal H}_{\rm{D2}}$ is often neglected,\cite{dyakonov86} a well
justified approximation for heterojunctions with a small Fermi
wave vector (${\cal H}_{\rm{D2}}$ was considered recently in a
different context).  \cite{miller03} However, we will show that
${\cal H}_{\rm{D2}}$ leads to two interesting effects  for small
few electron quantum dots. The first term in Eq.~(\ref{hd2fd})
with its hermitian conjugate couples the state
$|n_{+}n_{-}\sigma\rangle$ with
$|n_{+}+\sigma,n_{-}-2\sigma,-\sigma\rangle$ which are degenerate
at $\omega_{c}\approx \omega_{0}/\sqrt{2}$  (the exact location of
the anticrossing depends on Zeeman splitting).  The magnitude of
the anticrossing is given by
\begin{equation}
\Delta E\approx \frac{2^{3/4}}{\sqrt{3}}\frac{\gamma_{c}}{l_{0}^{3}}
\sqrt{n_{+}(n_{-}+1)(n_{-}+2)}.
\end{equation}
For a GaAs dot with $l_{0}=10$~nm, $\Delta E \sim 0.1$~meV, while
for GaSb, InAs, and InSb it can reach 1~meV. Note, however, that
this anticrossing appears only for $n_{+}+n_{-}>1$. It may have
interesting consequences for spin dependent transport through a
few electron QD. The third to sixth terms in Eq.~(\ref{hd2fd}) are
linear in $a_{\pm}$, $a^{\dag}_{\pm}$ leading to an enhancement of
the spin-orbit effect [Eq.~(\ref{vhat})]. This affects our
$g$-factor and $T_{1}$ calculations by as much as a factor of 2,
when $l_{0}\lesssim 10$~nm.

\begin{figure}
\includegraphics[width=3in]{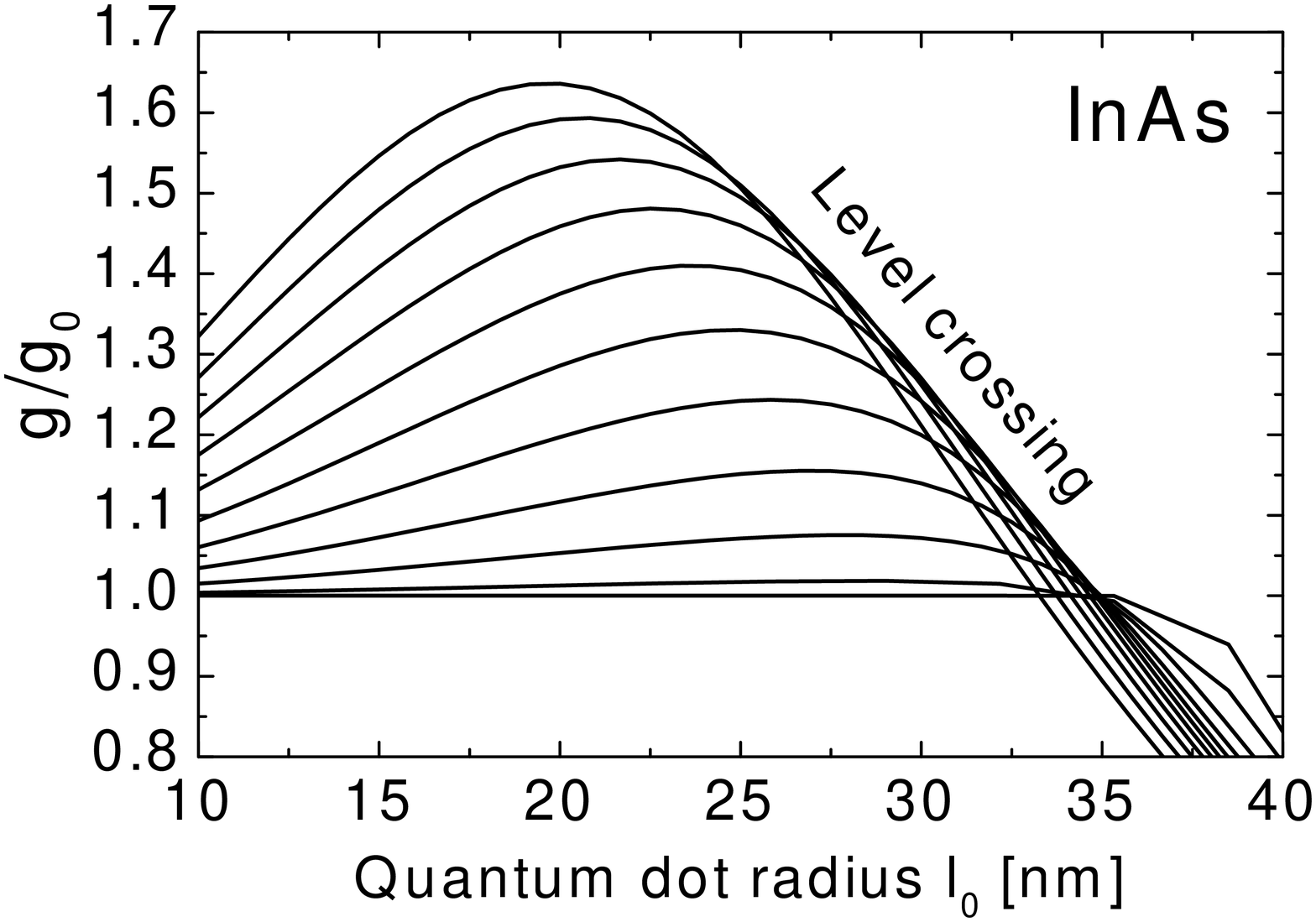}
\caption{Ratio between dot $g$-factor and bulk $g$-factor as a
function of the dot radius $l_{0}$ for an InAs heterojunction.
Each curve corresponds to an electric field, from bottom to top,
$E=10^{4}$, $1,2,\ldots, 10 \times 10^{5}$.  The magnetic field
applied in the growth [100] direction is assumed to be $1$~Tesla.
Notice the qualitative difference with respect to Fig.~1: Here
Rashba interaction dominates the $g$-factor, while in Fig.~1
Dresselhaus dominates. The results for InSb are similar.
\label{figthree}}
\end{figure}
%

Our QD ground state $g$-factor is defined by
$g=(E_{2}-E_{1})/(\mu_{B}B)$, $E_{1}$ and $E_{2}$ being the ground
and first excited states including spin. Considering
Eq.~(\ref{hd1hr}) as a second order perturbation to Eq.~(\ref{h0})
with $B$ field in the $[001]$ direction we get
\begin{eqnarray}
g &\approx& g_{0} + 2\frac{m_{e}m^{*}}{\hbar^{4}} \left[ 0.6
\gamma_{c}^{2}\kappa^{4}(1-\delta)-\alpha_{R}^{2}e^{2}E^{2}(1+\delta)\right]
\ell_{0}^{2}\nonumber\\
&&-\frac{1}{2}\frac{m_{e}m^{*3}}{\hbar^6}\left\{ 0.6
\gamma_{c}^{2}\kappa^{4}(1-\delta+\delta^2+\delta^3)\right.\nonumber\\
&&\left.-\alpha_{R}^{2}e^{2}E^{2}(1+\delta+\delta^2-\delta^3)\right\}
\omega_{c}^{2}\ell_{0}^{6}\nonumber\\
&&+ {\cal O}\left(\omega_{c}/ \omega_{0}\right)^{4}. \label{gpert}
\end{eqnarray}
Here $\delta=g_{0}m^*/m_e$, this expression being valid up to
second order in $\omega_{c}/\omega_{0}$ and the spin-orbit
admixtures. Clearly one sees that $g$-factor displays a rich
behavior as a function of QD radius and electric field. In
particular, $g-g_{0}$ will be positive and proportional to
$E^{4/3}$ if the Dresselhaus spin-orbit interaction is dominating
(GaAs and GaSb), but negative and proportional to $E^{2}$ when
Rashba dominates (InAs, InSb). In addition these effects increase
with increasing dot radius $l_{0}$, and there is a $B^{2}$
dependence at higher magnetic fields.

\begin{figure}
\includegraphics[width=3in]{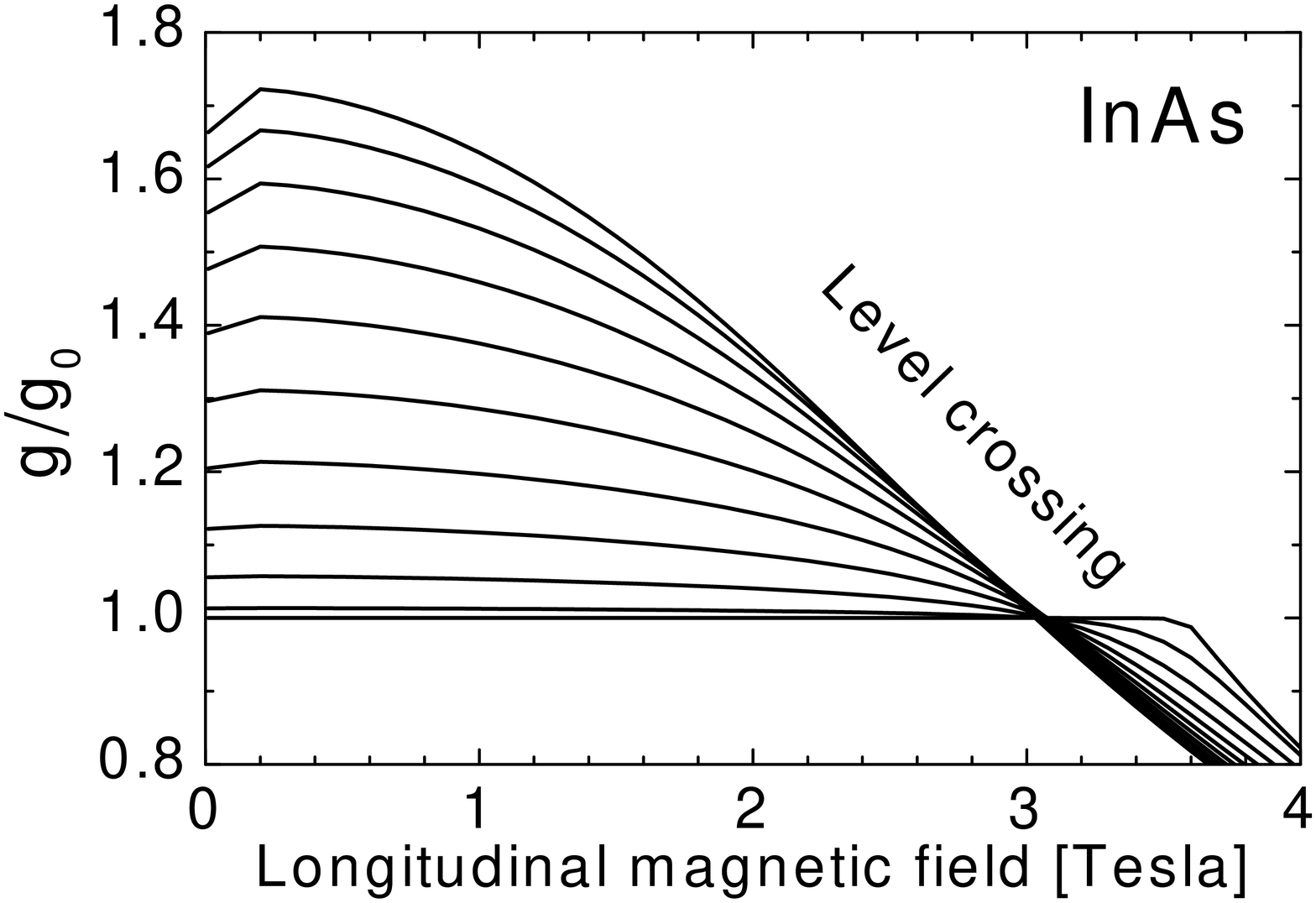}
\caption{$g/g_{0}$ as a function of magnetic field for an InAs
heterojunction quantum dot with $l_{0}=20$~nm. Electric fields are
the same as in Fig.~3. The results for InSb are similar.
\label{figfour}}
\end{figure}

We study the $g$-factor behavior at high electric and magnetic
fields, and large dot radius by resorting to exact diagonalization
of the full Hamiltonian [Eq.~(\ref{htotal})] (similar calculations
for Landau levels in a quantum well are available).\cite{lommer85}
Our basis consists of nine Fock-Darwin shells ($n_{+}+n_{-}\leq
8$), which together with spin leads to a $90\times 90$ matrix for
the Hamiltonian. Figs.~1 and 2 show results for GaAs, Figs.~3 and
4 for InAs. GaSb shows very similar behavior to GaAs, while InSb
is similar to InAs. The differences between the two sets of
materials is attributed to Dresselhaus spin-orbit interaction
dominating in Figs.~1 and 2, but Rashba dominating in Figs.~3 and
4 (the Rashba interaction appears to be dominant in SiGe
heterostructures, therefore $g$-factor behavior should be similar
to the InAs case considered here, except that $\alpha_{R}$ is
three orders of magnitude smaller and $g_{0}\approx 2$; hence the
effects considered here should be quite small for Si
heterojunctions: we estimate $g-g_{0}\sim -10^{-3}$ for $E\sim
10^{5}$~V/cm).\cite{wilamowski02} Our result suggests $g$-factor
can be controlled by a gate voltage (that either changes the
longitudinal electric field $E$ or the lateral confinement
$l_{0}$) \emph{as long as} $E \sim 10^{5}$~V/cm.  An important
feature of Figs.~1--4 is that a level crossing takes place for
large enough $l_{0}$ and $B$. In this regime the two lowest energy
states of the QD are approximate Landau levels with the same spin,
leading to extremely fast phonon emission rates (see below). Hence
a QD quantum computer should operate away from this level
crossing, which is actually a smooth anti-crossing for InAs (see
Figs.~3 and 4). Note that Figs.~1-4 plot the ratio of QD to bulk
$g$-factor ($g/g_{0}$), therefore the corresponding deviation in
QD Zeeman energy from the bulk value will be rather appreciable
for GaSb, InAs, and InSb since these materials have $g_{0}\sim
-10$ (Table~I).

We now turn to calculations of the transition rate between the two
lowest energy states  due to spontaneous phonon emission.  The
electron-piezophonon interaction\cite{khaetskii01}
\begin{equation}
{\cal U}_{\rm {e-ph}}^{q\alpha}= \sqrt{\frac{\hbar}{2\rho V
\omega_{q\alpha}}} e^{i \left({\mathbf q}\cdot {\mathbf r}
-\omega_{q\alpha}t\right)} eA_{q\alpha} b^{\dag}_{q\alpha}+
\rm{h.c.}, \label{eph}
\end{equation}
couples these states  in the presence of spin-orbit admixture.
Here $b^{\dag}_{q\alpha}$ creates an acoustic phonon with wave
vector ${\mathbf q}$ and polarization ${\bf\widehat{e}_{\alpha}}$
($\alpha = L, T_{1}, T_{2}$), $\rho$ is the material density, and
$V$ the volume of the sample. $A_{q\alpha}$ is the amplitude of
the electric field created by the phonon strain, which is given by
$\widehat{q}_{i}\widehat{q}_{k}e\beta_{ijk}e_{q\alpha}^{j}$, with
$\widehat{q}={\mathbf q}/q$, $e\beta_{ijk}=eh_{14}$ (see Table~I)
for $i\neq k$, $i\neq j$, and $j\neq k$. The polarization
directions are
\begin{eqnarray}
{\bf \widehat{e}}_{{\rm L}}&=&
(\sin{\theta}\cos{\phi},\sin{\theta}\sin{\phi},\cos{\theta}),
\\
{\bf \widehat{e}}_{{\rm T1}}&=&
(\cos{\theta}\cos{\phi},\cos{\theta}\sin{\phi},-\sin{\theta}),
\\
{\bf \widehat{e}}_{{\rm T2}}&=&
(-\sin{\phi},\cos{\phi},0).
\end{eqnarray}
The transition rate is given by
Fermi's golden rule,
\begin{equation}
\frac{1}{T_{1}}=\frac{V}{(2\pi)^{2}\hbar}\sum_{\alpha}\int
d^{3}q\left| \langle 1 |{\cal U}_{\rm{e-ph}}^{q\alpha} |2\rangle
\right|^{2}\delta (\hbar \omega_{q\alpha} - E_{2}+E_{1}),
\label{golden_rule}
\end{equation}
which under the same perturbative approximation as
Eq.~(\ref{gpert}) leads to
\begin{eqnarray}
\frac{1}{T_{1}}&\approx&
\frac{4}{105\pi}\left(\frac{1}{s_{T}^{5}}+\frac{3}{4}\frac{1}{s_{L}^{5}}\right)
\frac{4m^{*4}(eh_{14})^{2}}{\rho\hbar^{7}}\nonumber\\
&&\times\left(
0.61 \gamma_{c}^{2} \kappa^{4}
+\alpha_{R}^{2}e^{2}E^{2}\right)
\left(\frac{g_{0}\mu_{B}B}{\hbar}\right)^{5}
l_{0}^{8}
\nonumber\\
&&\times
(n_{ph}+1)
\left[1+{\cal O}\left(\omega_{c}/\omega_{0}\right)^{2}\right].
\label{t1pert}
\end{eqnarray}
Here $s_{T}$ and $s_{L}$ are the transverse and longitudinal
acoustic phonon velocities respectively. The spin-flip rate is
extremely sensitive to QD radius and external magnetic field. At
temperatures lower than Zeeman splitting, the emitted phonon
occupation number $n_{ph}$ is much smaller than one, and
Eq.~(\ref{t1pert}) is independent of temperature. At higher
temperatures Raman processes will dominate.\cite{khaetskii01}
\begin{figure}
\includegraphics[width=3in]{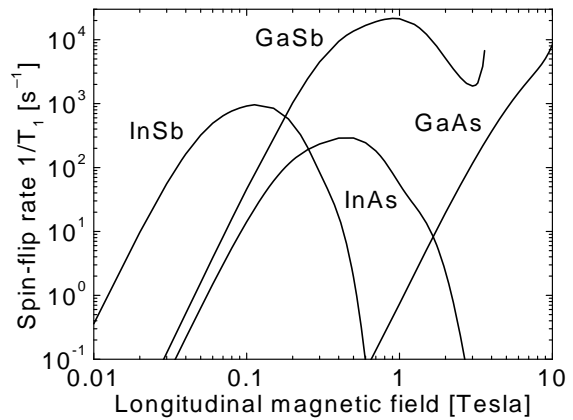}
\caption{Spin-lattice relaxation rate for the spin-doublet ground
state of a  20~nm one-electron quantum dot as a function of the
external magnetic field. Notice for all materials $1/T_{1}\propto
B^{5}$ at low $B$ fields. At high $B$ the Zeeman phonon wavelength
becomes smaller than the dot radius and the rate is strongly
suppressed. This effect is evident at $B\sim 1$~Tesla  for the
narrow gap materials, which have quite large bulk $g$-factors.
Here we assumed $E=10^{4}$~V/cm. The curves terminate at the level
crossing, when the rate is extremely enhanced because  the
transition states have the same spin. \label{figfive}}
\end{figure}
\begin{table}
\begin{tabular}{cc c c c c}
\hline\hline Parameter && GaAs & GaSb & InAs & InSb\\\hline
$g_{0}$      &                   & $-0.44$ & $-7.8$ & $-15$ & $-50.6$\\
$m^{*}/m_{e}$&                   & $0.067$ & $0.0412$ & $0.0239$ & $0.0136$\\
$\alpha_{R}$ & [\AA$^{2}$]       & $4.4$   & $33$   & $110$ &  $500$\\
$\gamma_{c}$ & [eV\AA$^{3}$]     & $26$    & $187$  & $130$ &  $228$\\
$eh_{14}$    & [$10^{-5}$ erg/cm]& $2.34$ & $1.5$ & $0.54$ & $0.75$\\
$s_{L}$    & [$10^{5}$ cm/s]& $5.14$ & $4.30$ & $4.20$ & $3.69$\\
$s_{T}$    & [$10^{5}$ cm/s]& $3.03$ & $2.49$ & $2.35$ & $2.29$\\
$\rho$    & [g/cm$^{3}$]& $5.3176$ & $5.6137$ & $5.6670$ & $5.7747$\\
\hline\hline
\end{tabular}
\caption{Parameters used in our calculations.\cite{silva97,cardona88}}
\end{table}

It is interesting to study deviations from the perturbative
approximation Eq.~(\ref{t1pert}). In particular, at high magnetic
fields the resonant phonon wavelength $\lambda_{Z}=hs/E_{Z}$
becomes much smaller than the dot radius making the  dipolar
approximation on the electron-phonon interaction inappropriate
[Eq.~(\ref{t1pert}) assumes the exponent in  Eq.~(\ref{eph}) can
be approximated by  $\sim 1+i{\mathbf q}\cdot {\mathbf r}$].
Furthermore, one immediately sees that $1/T_{1}$ is extremely
sensitive to the energy difference $E_{1}-E_{2}$, assumed equal to
the bulk Zeeman energy in Eq.~(\ref{t1pert}). Here we show
calculations of Eq.~(\ref{golden_rule}) using energy levels and
eigenstates obtained by exact diagonalization in a $90$
dimensional Fock-Darwin basis. In addition, we go beyond the
dipolar approximation  by using the identity
\begin{equation}
e^{i{\mathbf q}\cdot {\mathbf r}}= e^{-|\eta|^{2}}
e^{i\eta^{*}a_{+}^{\dag}}
e^{i\eta a_{+}}
e^{i\eta a_{-}^{\dag}}
e^{i\eta^{*}a_{-}},
\label{eiqr}
\end{equation}
where $\eta(\theta,\phi)=q\ell \sin{\theta}e^{i\phi}/2$ depends on
the polar angles of the phonon wave vector ${\mathbf q}$ [because
of this dependency, we have to perform the angular integrals in
Eq.~(\ref{golden_rule}) numerically].  Each of the exponents in
Eq.~(\ref{eiqr}) is expanded in powers of $\eta$, but we note that
within our subspace $n_{\pm}\leq 8$, therefore only up to the
ninth power needs to be retained. We checked the convergence of
our calculations by reducing the Fock-Darwin subspace and noting
that no appreciable change takes place for $n_{+}+n_{-}\geq 4$.
Our results agree with perturbation theory [Eq.~(\ref{t1pert})] at
low $B$ and $E$. Fig.~5 shows the spin-flip rate as a function of
the  magnetic field. It is evident that materials such as InAs and
InSb deviate from perturbation theory by more than three orders of
magnitude when $B$ is as low as  $1$~Tesla. This happens because
taking into account the full electron-phonon Hamiltonian leads to
an exponential decrease in the rate when $q\ell \gg 1$, since
Eq.~(\ref{eph}) oscillates appreciably in this regime. Note,
however, that $1/T_{1}\propto B^{5}$ at low enough $B$ for all
materials. Fig.~6 shows the dependency of the spin-flip rate with
lateral confinement radius $l_{0}$ in a GaAs QD. As the electric
field increases,  the dependency with $l_{0}$ displays an striking
behavior, which happens due to the sign change of $g$-factor shown
in Fig.~1. A small Zeeman energy implies negligible phonon density
of states, and hence the rate is zero for $l_{0}\sim 50$~nm and
$E=7\times 10^{5}$~V/cm in Fig.~6. We expect $1/T_{1}$ will behave
similarly to Fig.~6 when $g$ changes sign due to barrier
penetration in AlGaAs,\cite{salis01} this property being extremely
useful in the initialization and decoherence suppression of a QD
quantum computer.

We now discuss possible corrections to the simple model discussed
here. At strong confinement in the 100 direction  one expects
$\Gamma$-X valley mixing to become important. For GaAs, $E_{\Gamma
X}=0.48$~eV, which is comparable to $E_{0z}$ [Eq.~(\ref{groundz})]
only when the electric field is the highest considered here, $E >
10^{6}$~V/cm. This also holds true for InAs and InSb, but in GaSb
$\Gamma$-X coupling will be important for $E
> 5\times 10^{5}$~V/cm. Therefore even though a full ${\mathbf
k}\cdot {\mathbf p}$ calculation would yield some
corrections,\cite{kiselev98}  we do not expect it to change our
results qualitatively in the range considered here. The same holds
true for the inadequacy of the Rashba Hamiltonian,  which starts
deviating from Eq.~(\ref{hrashba}) when $E\sim
10^{6}$~V/cm.\cite{silva97}
%

%
\begin{figure}
\includegraphics[width=3in]{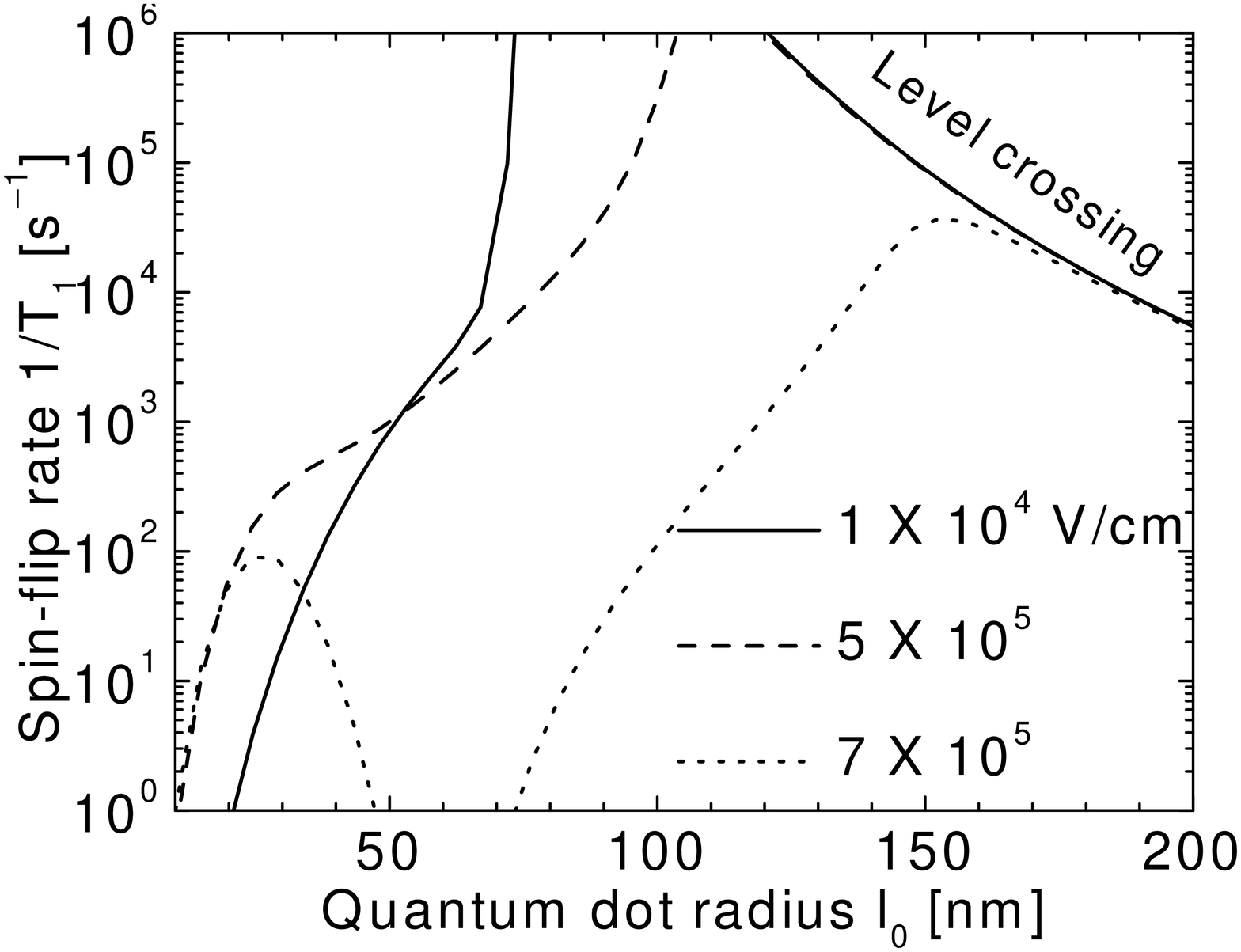}
\caption{Spin-flip rate $1/T_{1}$ due to spin-orbit admixture as a
function of dot radius in GaAs. For small radius, $1/T_{1}\propto
l_{0}^{8}$. For $E=7\times 10^{5}$~V/cm there is a striking change
in behavior: This happens due to the sign change in $g$-factor
seen in Fig.~1. \label{figsix}}
\end{figure}

Before concluding, we discuss the approximations and the
limitations of this work. The most essential approximation of our
model, the use of the ${\bf k}\cdot {\bf p}$ perturbation theory
within an effective mass approximation scheme to describe the
conduction band, has been extensively used in the
literature\cite{lommer85} and should be well-valid for the problem
we study. We have made two additional non-essential approximations
in our theory in order to simplify our numerical computations: The
triangular well approximation for the $z$-confinement of the wave
function and the parabolic well Fock-Darwin confinement
approximation in the 2D $x-y$ plane. These approximations are
reasonable enabling us to produce numerical results for a range of
system parameters in several different semiconductor structures,
which would have been difficult, if not impossible, to carry out
had we used more realistic (and thereby numerically more
demanding) quantum dot confinement potentials (The triangular well
approximation was employed recently to derive new results
regarding D'yakonov-Perel' relaxation anisotropy for conduction
electron spins confined by heterojunctions with $E\gtrsim
10^5$~V/cm).\cite{averkiev02} Our most important qualitative
result, establishing the viability of controlling the spin
dynamics of III--V semiconductor quantum dots (both $g$-factor and
$T_{1}$ engineering) by using external gates to suitably
manipulate the spin-orbit coupling through Dresselhaus and Rashba
effects, should be completely independent of these approximations.
In fact, we expect that the use of more sophisticated confinement
models may actually make the gate control effects we predict
somewhat stronger by pushing the required electric fields (and
consequently 2D carrier densities) to somewhat lower values than
our predicted $10^{5}$~V/cm range. The main limitation of our
predicted spin-orbit coupling induced gate control effect is, in
fact, the rather large electric fields ($\sim 10^{5}$~V/cm) and
the associated 2D carrier densities ($\sim 10^{12}$~cm$^{-2}$)
that are required to produce significant gate control effects.

In conclusion, we show that quantum dot longitudinal $g$-factor
and spin-flip time $T_{1}$ can be controlled electrically even in
the absence of wave function overlap with a different material.
These parameters show a striking dependence with dot radius and
magnetic field when the 2DEG confinement is strong (electric field
$E\sim 10^{5}$~V/cm). For example, the $g$-factor changes sign and
$T_{1}$ is extremely sensitive to the dot radius.  $g$-factors for
one-electron dots can be measured using transport
spectroscopy.\cite{hanson03} We show that $T_{1}$ is drastically
increased in narrow gap materials (InAs, InSb) due to deviations
from the dipolar approximation in the electron-phonon interaction,
suggesting these materials are promising for the fabrication of a
quantum dot spin quantum computer.  $T_{1}$ is found higher than
$10^{-4}$~s under quite different circumstances (see Figs.~5 and
6) showing that small III--V one-electron quantum dots
($l_{0}<50$~nm) will have their low temperature phase coherence
time ($T_{2}$) dominated by nuclear induced spectral
diffusion.\cite{desousa03} This result establishes the versatility
of III--V quantum dots as units for single spin manipulation.  A
related finding of interest in our work is the dual importance of
both Dresselhaus (i.e. the bulk inversion asymmetry inherent in
Zincblende structures of III--V semiconductors) and Rashba (i.e.
the real space structural inversion asymmetry present in a
heterostructure due to external electric fields) spin-orbit
coupling terms in semiconductor nanostructures -- in particular,
for GaAs and GaSb quantum dot structures investigated in this
work, we typically find the bulk inversion asymmetry (i.e.
Dresselhaus) effect to be quantitatively more important than the
Rashba effect. The relative quantitative importance of the
Dresselhaus effect in III--V nanostructures should have
considerable significance not only in the $g$-factor engineering
and the spin relaxation time control of relevance to the spin
quantum computer architecture (that we consider in this work), but
also in the fabrication of the Datta-Das spintronic
transistor\cite{datta90} where spin-orbit coupling is used to
modulate a spin-polarized current in a field effect transistor
configuration.  The authors acknowledge discussions with A.
Kaminski and I.  \v{Z}uti\'{c}.  This work is supported by ARDA,
LPS, US-ONR, and NSF.


\begin{thebibliography}{99}

\bibitem{dassarma01} S. Das Sarma, J. Fabian, X. Hu, and
I. \v{Z}uti\'{c}, Solid State Comm. {\bf 119}, 207 (2001).

\bibitem{recher00} P. Recher, E.V. Sukhorukov, and D. Loss, \prl {\bf
  85}, 1962 (2000).

\bibitem{zutic02} I. \v{Z}uti\'{c}, J. Fabian, and S. Das Sarma,
\prl {\bf 88}, 066603 (2002).

\bibitem{loss98} D. Loss and D.P. DiVincenzo, Phys. Rev. A {\bf 57},
120 (1998); G. Burkard, D. Loss, and D.P. DiVincenzo, \prb {\bf 59},
2070 (1999); X. Hu and S. Das Sarma, Phys. Rev. A {\bf 61}, 062301
(2000); M. Friesen, P. Rugheimer, D.E. Savage, M.G. Lagally, D.W. van
der Weide, R. Joynt, M.A. Eriksson, \prb {\bf 67}, 121301 (2003).

\bibitem{datta90} S. Datta and B. Das, Appl. Phys. Lett. {\bf 56}, 665 (1990).

\bibitem{nitta97} J. Nitta, T. Akazaki, H. Takayanagi, and T. Enoki, \prl
{\bf 78}, 1335 (1997); T. Koga, J. Nitta, T. Akazaki, and
H. Takayanagi, \prl {\bf 89}, 046801 (2002).

\bibitem{bychkov84} Y.A. Bychkov and E.I. Rashba, J. Phys. C {\bf 17},
6039 (1984).

\bibitem{silva97} E.A. de Andrada e Silva, G.C. La Rocca, and
F. Bassani, \prb {\bf 55}, 16293 (1997); \textit{ibid.} {\bf 50}, 8523 (1994).

\bibitem{dresselhaus55} G. Dresselhaus, Phys. Rev. {\bf 100}, 580 (1955);
M. D'yakonov and V.I. Perel', Sov. Phys. - Solid State, {\bf 13}, 3023 (1972).

\bibitem{tarucha96} S. Tarucha, D.G. Austing, T. Honda, R.J. van der
Hage, and L.P. Kouwenhoven, \prl {\bf 77}, 3613 (1996).

\bibitem{jacak98} L. Jacak, A. W\'{o}js, and P. Hawrylak, {\it Quantum
  Dots} (Springer Verlag, Berlin, 1998).

\bibitem{khaetskii01} A.V. Khaetskii and Yu.V. Nazarov, Phys. Rev. B
{\bf 64}, 125316 (2001).

\bibitem{salis01} G. Salis, Y. Kato, K. Ensslin, D.C. Driscoll, A.C.
Gossard, and D.D. Awschalom, Nature {\bf 414}, 619 (2001); Y. Kato,
R.C. Myers, D.C. Driscoll, A.C. Gossard, J. Levy, and D.D. Awschalom,
Science {\bf 299}, 1201 (2003).

\bibitem{jiang01} H. W. Jiang and E. Yablonovitch, \prb {\bf 64},
041307 (2001).

\bibitem{woods02} L.M. Woods, T.L. Reinecke, and Y. Lyanda-Geller,
\prb {\bf 66}, 161318 (2002).

\bibitem{vrijen00} R. Vrijen, E. Yablonovitch, K. Wang, H.W. Jiang, A.
Balandin, V. Roychowdhury, T. Mor, and D. DiVincenzo, \pra {\bf 62},
012306 (2000).

\bibitem{dyakonov86} M.I. D'yakonov and V.Yu. Kachorovskii,
Sov. Phys. Semicond. {\bf 20}, 110 (1986).

\bibitem{schliemann03} J. Schliemann, J.C. Egues, and D. Loss,
\prl {\bf 90}, 146801 (2003).

\bibitem{miller03} J.B. Miller, D.M. Zumbuhl, C.M. Marcus,
Y.B. Lyanda-Geller, D. Goldhaber-Gordon, K. Campman, A. C. Gossard,
\prl {\bf 90}, 076807 (2003).

\bibitem{lommer85} G. Lommer, F. Malcher, and U. R\"{o}ssler, \prb
{\bf 32}, 6965 (1985); Supperlattices and Microstructures {\bf 2}, 273
(1986); E.L. Ivchenko, A.A. Kiselev, and M. Willander, Solid State
Commun. {\bf 102}, 375 (1997).

\bibitem{wilamowski02} Z. Wilamowski, W. Jantsch, H. Malissa, and U.
R\"{o}ssler, \prb {\bf 66}, 195315 (2002); Z. Wilamowski, W. Jantsch,
N. Sandersfeld, M. Muhlberger, F. Schaffler, and S. Lyon, Physica E
{\bf 16}, 111 (2003).

\bibitem{cardona88} M. Cardona, N.E. Christensen, and G. Fasol, \prb
{\bf 38}, 1806 (1988).

\bibitem{kiselev98} A.A. Kiselev, E.L. Ivchenko, and U. R\"{o}ssler,
Phys. Rev. B {\bf 58}, 16353 (1998).

\bibitem{averkiev02} N.S. Averkiev, L.E. Golub, and M. Willander,
Semiconductors {\bf 36}, 91 (2002).

\bibitem{hanson03} R. Hanson, B. Witkamp, L.M.K. Vandersypen,
L.H.W. van Beveren, J.M. Elzerman, and L.P. Kouwenhoven,
cond-mat/0303139 (unpublished).

\bibitem{desousa03} R. de Sousa and S. Das Sarma, Phys. Rev. B
{\bf 67}, 033301 (2003); R. de Sousa and S. Das Sarma, cond-mat/0211567
(unpublished).

\end{thebibliography}
\end{document}